# Exploring the Relation between NPP-VIIRS Nighttime Lights and Carbon Footprint, Population Growth, and Energy Consumption in the UAE

Fahim Abdul Gafoor[1], Chung Suk Cho[1] and Maryam R. Al Shehhi[1,*]

[1] Civil, infrastructural and Environmental Engineering, Khalifa University, Abu Dhabi, UAE

* Correspondence: maryamr.alshehhi@ku.ac.ae; Tel.: +971 2 312 4260

**Abstract:** Due to global warming and its detrimental effect, every country is responsible to join the global effort to reduce carbon emissions. In order to improve the mitigation plan of climate change, accurate estimates of carbon emissions, population, and electricity consumption are critical. Carbon footprint is significantly linked to the socioeconomic development of the country which can be reflected in the city's infrastructure and urbanization. We may be able to estimate the carbon footprint, population growth, and electricity consumption of a city by observing the nighttime light reflecting its urbanization. This is more challenging in oil-producing countries where urbanization can be more complicated. In this study, we are therefore investigating the possibility of correlating the remotely sensed NPP-VIIRS Nighttime light (NTL) estimation with the aforementioned socioeconomic indicators. Daily NPP-VIIRS NTL were obtained for the period between 2012 to 2021 for the United Arab Emirates (UAE) which is one of the top oil producing countries. The socioeconomic indicators of the UAE, including the population, electricity consumption, and carbon dioxide emissions, have been obtained for the same period. The analysis of the correlation between the NTLs and the population indicates that there is a high correlation of more than 0.9. There is also a very good correlation of 0.7 between NTLs and carbon emissions and electricity consumption. However, these correlations differ from one city to another. For example, Dubai has shown the highest correlation between population and NTLs ($R^2 > 0.8$). However, the correlation was the lowest in Al-Ain, a rural city ($R^2 < 0.4$) with maximum electricity consumption of $1.1 \times 10^4$ GWh. These results demonstrate that NTLs can be considered as a promising proxy for carbon footprint and urbanization in oil-producing regions.

## 1. Introduction

Many countries do not have access to accurate measurements of carbon footprints and electricity consumption due to the expensive instrumentation, poor facilities, and confidentiality issues [1–3]. Thus, the data especially in the poor countries may not be accurate or are not consistently reported. For instance, there is not enough data for third-world countries, such as the sub-Saharan countries [4]. In parallel, countries are obliged to reduce their carbon dioxide emissions ($CO_2$) by 7.6% each year to reduce the climate changes symptoms and keep temperatures from rising beyond 1.5 degrees Celsius

[5]. However, unless the $CO_2$ emissions and energy consumption are accurately estimated, this cannot be achieved.

As a result, efforts have been made to quantify the carbon footprint and electricity consumption. One of the most reliable approaches available is the ISO 14040 family of standards which is commonly used to determine the greenhouse gas (GHG) emissions of a product during its life cycle. According to ISO 14040 to estimate the carbon footprint, the direct atmospheric emissions from the operation of the facilities included within organizational boundaries are required [6] which are obtained from the available databases. Another approach of developing a consumption-based GHG inventory was developed based on the data collected from Census block groups, cities, and counties. The basic approach consists of calculating average household carbon footprints and then creating population-weighted averages for each city, state, and the county as a whole. The average household's consumption of energy, transportation fuels, water, waste, building and construction materials, goods, and services is estimated, multiplied by GHG emission factors, and summed up [7]. In another study, the household carbon footprint of multiple cities is estimated using an input–output approach together with mixed-unit consumption data using the lifestyle data and physical consumption of goods, leisure, and services and monetary expenditure [8].

Considering that most carbon footprints are produced by the combustion of fossil fuels for energy generation and industrial development, recent studies have examined the use of satellite technology to determine industrialization and carbon footprint based on nighttime remotely sensed data. The monthly statistical data of electric power consumption (EPC) for example in 14 provinces of southern China have been quantitatively correlated with the monthly composite of the nighttime light (NTL) imagery acquired from the Visible Infrared Imaging Radiometer Suite (VIIRS) Day/Night Band (DNB) data [9]. The study successfully obtained high coefficients of determination ($R^2$) of 0.8816 and 0.87 using both polynomial and linear regressions. The study also reported that these correlations were derived using the monthly data unlike the previous studies which have primarily examined the quantitative relationship between NTLs and statistical variables over relatively long-time scales (especially one year). Shi et al. (2014) have also found that the linear regression shows that $R^2$ values of NTL from NPP-VIIRS are high, can reach up to 0.8, when correlated with Gross Domestic Product (GDP) and EPC at multiple scales in China [10]. The VIIRS-NPP NTLs have been also used as a proxy to accurately measure $CO_2$ emissions at the city scale in China [11]. They could successfully develop a linear regression model between the estimated value and the statistical value of carbon dioxide with the $R^2$ around 0.85.

According to previous studies, such as those reported here, there is a correlation between the NTL and socioeconomic variables, but these studies do not examine the long-term changes in the NTL nor their variability with the urbanization structure. Thus, the aims of this study are to: 1) find the correlation between carbon emissions (carbon footprint), electricity consumption, population and NTLs 2) study trends in NTLs of VIIRS-NPP over a period of ten years, 3) determine urbanization's variability based on NTLs. The study area selected is the United Arab Emirates (UAE), which is one of the world's largest oil producers and the country has experienced significant civil development in the last three decades. As such, it is a vital case study to examine the eligibility for using the VIIRS-NPP to study its socioeconomics and urbanization.

## 2. Data and Methods

### 2.1. Study area: the UAE

With a total GDP of $420 billion in 2019, the UAE has benefited greatly from its oil exports since the 1960s. The country is largely dependent on fossil fuels and natural gas for its operation [12]. In addition, during the period 1950-2010, the UAE's urban population increased from 54.4% to 84.4% (Figure 1a). Thus, UAE has one of the highest rates of urbanization in the world, reaching 2.9% during the 2005-2010 period [13]. This has resulted in the UAE being one of the high carbon footprint countries, as each person produces 23.37 tons of $CO_2$ emissions per year according to the world bank database (Figure 1b), in addition to the high electricity consumption (Figure 1c). There are 16% of $CO_2$ emissions in the UAE generated by transportation, 44% by other industrial combustion and 32% by power [14]. The UAE, however, recently published national plans, such as Abu Dhabi 2030, to reduce the carbon footprint to a certain level [15].

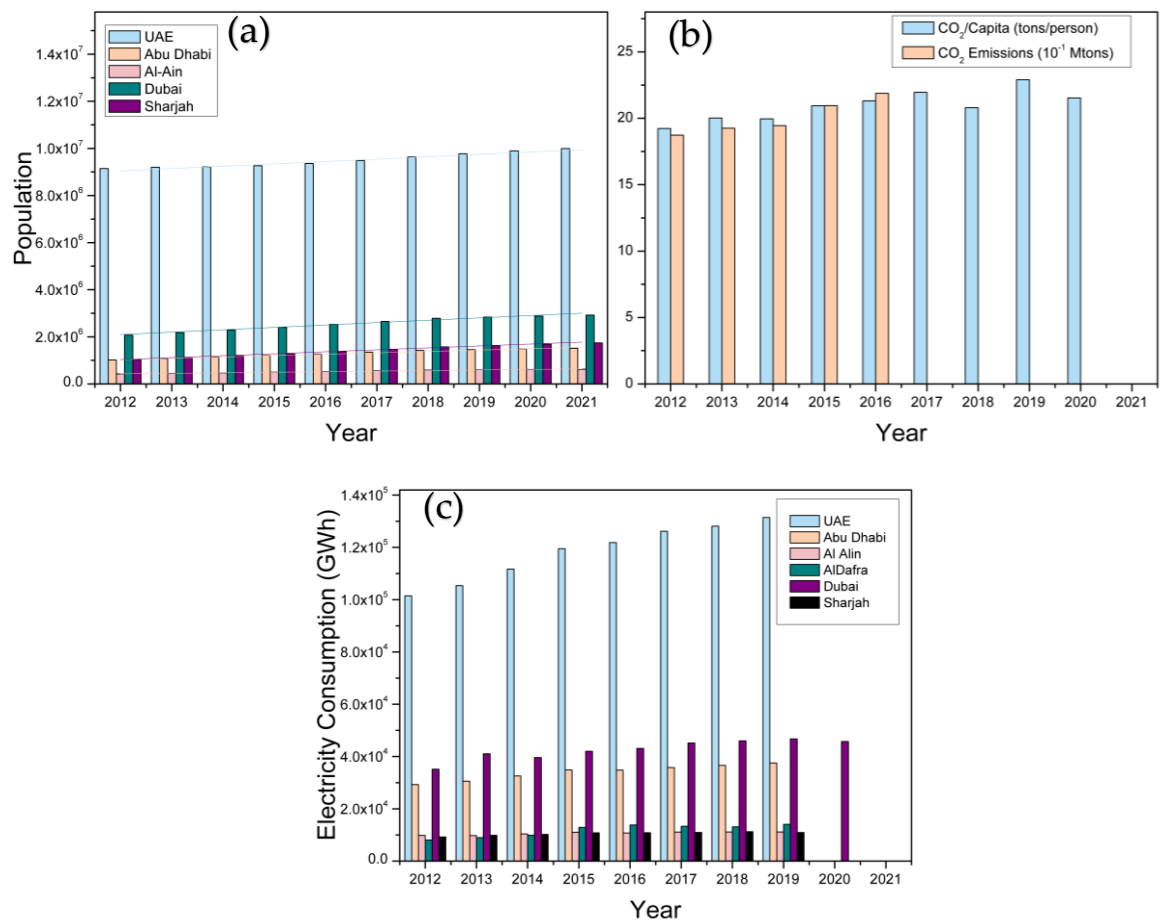

**Figure 1.** Changes of (a) population, (b) CO2 and CO2 per capita in the UAE between 2012 and 2021 and (c) Electricity Consumption in the UAE and in four individual cities between 2012 and 2021.

In addition, the UAE's recent initiative 'Net Zero by 2050' is a national campaign to achieve net-zero emissions by 2050, making the UAE to become the first Middle East and North Africa (MENA) nation to achieve this ambitious goal. As a consequence, the country starts to record and monitor all emissions and indicators to cope with climate change and establish itself as one of the forefront countries that supports the transition to renewable energy. Figure 2a shows the map of the UAE which consists of different cities and sub-cities that will be discussed and analyzed in this work. The figure also shows the UAE's urbanization structure including roads, coastal area and natural regions.

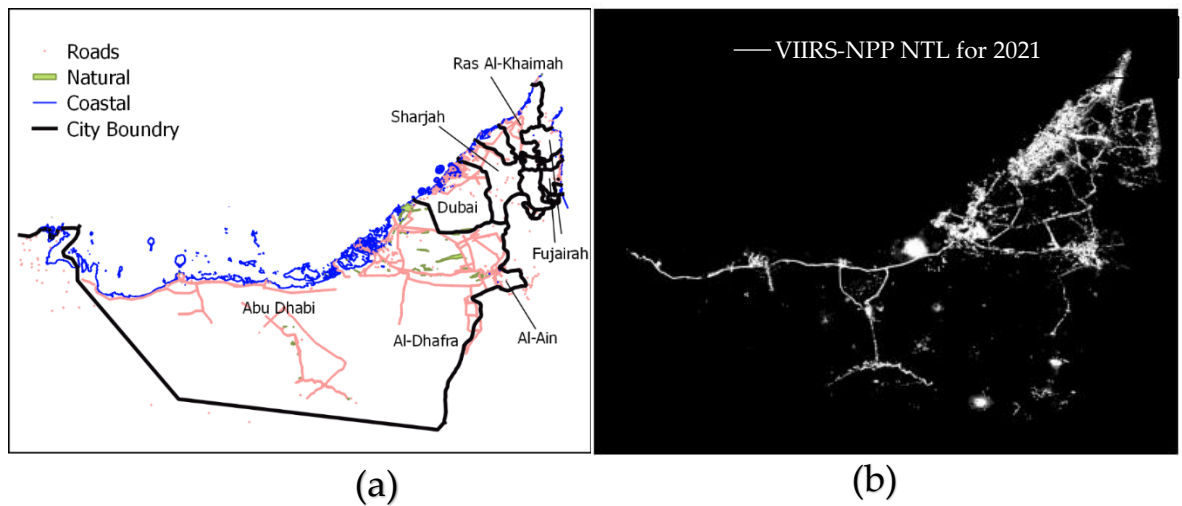

**Figure 2.** a) The map of the UAE showing the coastal line in blue, land boarder in black and b) the VIIRS-NPP Nighttime light over the UAE for the year 2021.

*2.2. Data and Satellite imageries*

The annual EC data [GWh] for the UAE and five different cities have been obtained for the years between 2012 and 2020. Additionally, carbon emissions (Mtons) and carbon emissions per capita have also been collected for the entire UAE for the same period. Furthermore, the population has been obtained for the entire UAE and for four specific cities (Abu Dhabi, Sharjah, Al-Ain, and Dubai). All of these data have been obtained from locally published reports by the UAE governmental entities such as Dubai Electricity & Water Authority (DEWA) and Abu Dhabi Water & Electricity Authority (ADWEA).

As for the satellite data, the Day-Night Band (DNB) sensor on board of the Visible Infrared Imaging Radiometer Suite (VIIRS) of the Suomi National Polar-orbiting Partnership (SNPP) is used. In particular, VIIRS/NPP Daily Gridded Nighttime light data (product: VNP46A1) have been retrieved with a 500-meter resolution. The VNP46A1 product contains 26 Science Data Sets (SDS) that include sensor radiance, zenith and azimuth angles (at-sensor, solar, and lunar), cloud-mask flags, time, shortwave IR radiance, brightness temperatures, VIIRS quality flags, moon phase angle, and moon illumination fraction. To ensure a high level of image quality, the moon illumination fraction threshold was set at 50% which has been found to improve the quality of the retrieved NTL.

Images are also discarded if solar zenith angles are less than 101°. The cloud mask has been applied afterwards to remove the cloudy days. As for the High Energy Particle (HEP), an initial filtering is done using thresholds placed on the DNB radiance and spike height index (SHI) [16]. SHI is calculated in two stages. First, by computing the relative difference between the pixel's radiance and the average of the two adjacent pixels. This calculation is performed for the two adjacent pixels in the same scan line (or row), and then for the two adjacent pixels in the same sample position (or column). The final SHI is the minimum of these two intermediate values. The resultant pixels with SHI values greater than 0.995 that also have radiance values greater than 1000 nW are filtered out. This eliminates the brightest of the HEP detections. Flaring sites have also been detected for grid cells with average temperature exceeding 1200 K and 1 % frequency. The outliers have been also removed as the primary purpose of the outlier removal is to exclude pixels with biomass burning, which are expressed as anomalously high radiance pixels but occur infrequently over an entire year. The high side outlier removal has the additional benefit of filtering out some of the aurora affected pixels. Filtering out the low radiance outliers has more subtle effect on the average, by filtering out observations that may be affected by clouds (errors in the cloud mask) or power outages.

NTL data were gathered for the UAE over the period between 2012 and 2021, and monthly and annual means were then calculated for the entire UAE and for the above-mentioned cities and urbanization zones including farms, industrial and urban areas. In addition, the averaged summer and winter NTLs for the cities have been also obtained for the seasonal analysis. Figure 2b shows the average VIIRS-NPP Nighttime light over the UAE for the year of 2021.

*2.3. Correlations and trend analysis*

2.3.1. Correlation and Regression

The correlation analysis was used in this study to determine the strength of the linear relationships between the NTLs and the socioeconomic variables, the EC, carbon emissions, and the population, and to compute their association. In addition, the ordinary least square (OLS) regression was used to model the relationship between the socioeconomic variables' response and the NTLs for the UAE and aforementioned cities. The general formula of the linear regression is shown in Eq.1. In order to evaluate the goodness of fit of the regression model, the coefficient of determination ($R^2$) is used which is described in Eq. 2.

$$y_i = \beta_o + \beta_i x_i \quad (1)$$

Where $y_i$ refers to the socioeconomic variables' response and $x_i$ refers to the NTLs. $\beta_o$ and $\beta_1$ are the coefficient of the linear OLS model.

$$R^2 = 1 - \frac{\sum(y_{i,\text{obs}} - y_{i,\text{est}})^2}{\sum(y_{i,\text{obs}} - \overline{y_{i,\text{obs}}})^2} \quad (2)$$

Where $y_{i,\text{obs}}$ is the actual value and $y_{i,\text{est}}$ is the estimated value.

2.3.2. Trend Analysis: Mann-Kendall test and Sen's slope estimator

In this study, the trends of the NTLs were estimated using a non-parametric Mann-Kendall test and Sen's slope which works for all distributions [17–19]. The Kendall's tau-b ($T_b$) correlation coefficient and Sen's slope can be obtained by the steps illustrated in Table 1.

Table 1. Mann-Kendall test and Sen's slope approach: detailed test statistics.

| Statistical test | Formula | |
|---|---|---|
| Mann-Kendall | (1) $S = \sum_{k=1}^{n-1} \sum_{j=k+1}^{n} sgn(x_j - x_k)$ | where n is the length of the sample, $x_j$ and $x_k$ are from $k=1, 2, …, n-1$ and $j= k+1, …, n$. If n is bigger than 8, statistic S approximates to normal distribution |
| | (2) $sgn(x_j - x_k) = \begin{cases} +1, if(x_j - x_k) > 0 \\ 0, if(x_j - x_k) = 0 \\ -1, if(x_j - x_k) < 0 \end{cases}$ | |
| | (3) $var(S) = \frac{n(n-1)(2n+5)}{18}$ | var(S) is the variance of S |
| | (4) $T_b = \begin{cases} \frac{S-1}{\sqrt{var(S)}}, if\ S > 0 \\ 0, \quad if\ S = 0 \\ \frac{S+1}{\sqrt{var(S)}}, if\ S < 0 \end{cases}$ | $T_b$ is the test statistic. If $T_b>0$, it indicates an increasing trend, and vice versa |
| Sen's slope | $\beta = Median\left(\frac{x_j - x_i}{j - i}\right),\ j > i$ | $\beta>0$ indicates upward trend in a time series. Otherwise, the data series presents downward trend during the time period |

## 3. Results and Discussions

### 3.1. Analysis of CO₂, electricity and population

A significant increase in the population of the UAE has occurred over the past decade, owing to rapid economic growth in the country and the establishment of international investment projects that have opened up new job markets in the country. Hence, its population has increased from around 9 million in 2012 to around 9.9 million in 2021. This means that the population increased by 11%, and the growth is expected to continue in the future. Of these 9.9 million, 8.84 million were expatriates, which constituted approximately 89% of the entire population. Migrants from India and Pakistan combined were 4.02 million, constituting the highest percentage of expatriates (i.e., 46% of the expat population) [18]. According to a comparison of four cities in the UAE, Dubai has the highest population density, which is about 22% of the total population. However, the capital city, Abu Dhabi, is ranked second in terms of population, followed by Sharjah and Al-Ain. The relative population density over the period from 2012 to 2021 remains the same in terms of the ranking, although there is a general upward trend across all the cities, which is particularly noticeable in the city of Dubai.

The increase in population has led to a significant increase in electricity consumption, as depicted in Figure 1a, c. This is due to the rapid economic growth, financial development, and urbanization which have influenced the consumption of electricity in various ways. For example, economic growth may increase the purchasing power of households for energy-efficient appliances, which may result in higher electricity consumption [13]. The resultant increase in the electricity consumption is ranged from $1 \times 10^{-5}$ GWh to $1.3 \times 10^{-5}$ GWh for the period between 2012 and 2019. However, a slight decline was observed in 2018 and 2020 as indicated by the data. In comparison to other cities, Dubai has the highest power consumption, which equals to 35% of total consumption. Abu Dhabi is ranked second in electricity consumption, which was around $3.4 \times 10^4$ GWh in 2012 and increased to $3.7 \times 10^4$ GWh in 2019. However, electricity consumption is less in Sharjah, Al-Ain, and Al-Dhafra and it is approximately $1 \times 10^4$ GWh. Among these cities, the demand for electricity in Al-Dhafra has increased significantly due to the expansion of the region and increased population especially between 2012 and 2017.

With the increase in the population, the total $CO_2$ emissions have shown an increasing trend as well between 2012 to 2021. However, slight decrease in 2018 and 2020 as observed in the data. In addition, the carbon footprint per capita has increased from 18 to 22 tons $CO_2$ per capita in the UAE between 2012 to 2016 (Figure 1b). The rapid increase of expat migrants who are mostly working-age population actively engage in the labor market, also contributed to producing a significant carbon footprint.

*3.2. Temporal and spatial analysis of NTLs*

The daily NPP-VIIRS nighttime light have been extracted for six different cities in the UAE for the period between 2011 to 2021 as shown in Figure 3. Among the six cities, the city of Dubai has shown the steepest increase in the NTL from $2 \times 10^5$ in 2011 to $4 \times 10^5$ in 2021. This means that the NTL values have almost doubled in this period. Sharjah is found to be the second city after Dubai in the significant increase in the NTL from $2 \times 10^5$ in 2011 to $3 \times 10^5$ in 2021. This is followed by Fujairah, Ras Al Khaimah and Al-Ain which show much lower total NTLs of $5.4 \times 10^4$ in 2011 and $7 \times 10^4$ in 2021 in the city of Fujairah and $6 \times 10^4$ in 2011 to $8 \times 10^4$ in 2020 in Ras Al Khaimah. The NTL values at these cities reflect the low urbanization development in the cities given their governing rural structure and the mountainous region rather than urban cities.

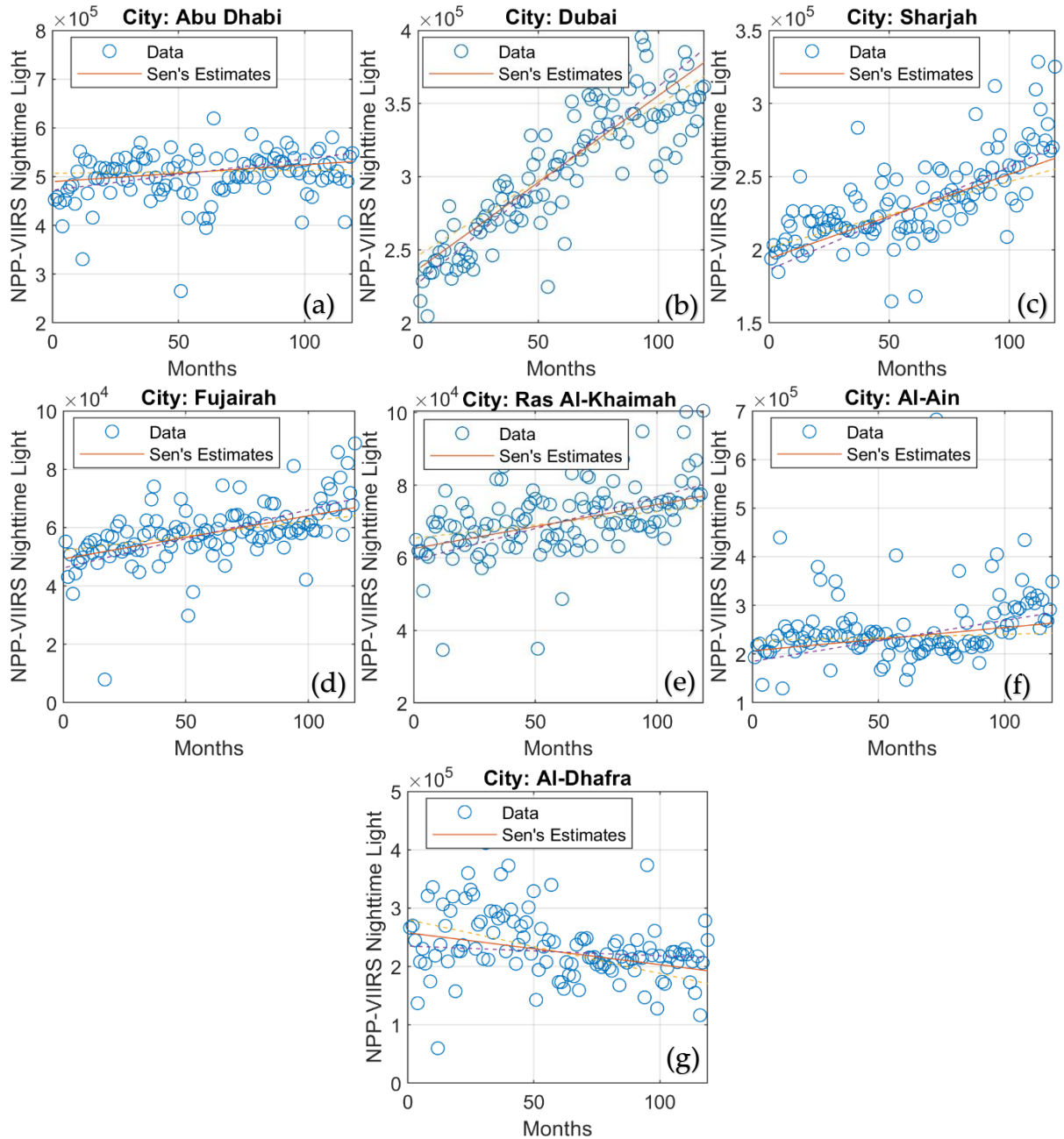

**Figure 3.** NPP-VIIRS Night time lights and estimated Sen's Slope of NPP-VIIRS Nighttime Light versus the months between 2011 and 2021 for different cities in the UAE: (a) Abu Dhabi, (b) Dubai, (c) Sharjah, (d) Fujairah, (e) Ras Al-Khaimah, (f) Al-Ain and (g) Al-Dhafra.

In contrast, NTLs have not shown noticeable increasing trend in the cities of Abu Dhabi and Al-Dhafra due to the prior developed infrastructure of the two cities and minor urbanization changes have been obtained in the last decade unlike the other cities which have undergone significant infrastructural development. This could be also observed by the high NTL values in Abu Dhabi compared to the other cities, including Dubai, reaching a value of $6\times10^5$.

To quantify the trends of the NTL changes over the decade, Mann-Kendall test and Sen's slope were obtained for six cities as shown in Figure 3. The $T_b$ which represents the degree of the trend is also shown in Table 2 along with the significance level (*P*-value). Dubai has shown the highest positive (i.e., increasing) trend with $T_b$ of 0.7 (*P*-value <0.05) followed by Sharjah ($T_b$=0.6), Fujairah ($T_b$=0.5), Ras Al Khaimah ($T_b$ = 0.4) and Al-Ain ($T_b$= 0.2). However, as expected low $T_b$ values were obtained for the cities of Abu Dhabi and Al-Dhafra ($T_b$<0.2). Similarly, Sen's slope shown in Figure 3 can prove that Dubai shows the highest increasing trend of NTL followed by the cities mentioned earlier. In addition, Sen's slope is found to be flat for the city of Abu Dhabi confirming the weak increase in the NTL in the city.

**Table 2.** Man-Kendall Analysis for the Nighttime light over different cities between 2012 and 2021.

| Region | $T_b$ | *P-value* |
|---|---|---|
| Al-Dhafra | -0.23601 | 1.43E-04 |
| Abu Dhabi | 0.1967 | 0.001527 |
| Al-Ain | 0.224 | 2.95E-04 |
| Ras Al-Khaimah | 0.380145 | 8.91E-10 |
| Fujairah | 0.462755 | 8.66E-14 |
| Sharjah | 0.553625 | 4.45E-19 |
| Dubai | 0.697194 | 2.58E-29 |

These trends are consistent with the spatial trend map generated for the UAE as shown in Figure 4. The positive values show increasing trend whereas the blue values show decreasing trend. As shown in the map Dubai show significant increase in the NTL with the spatial trend more than 2 (red color). In addition, the infrastructural development projects could be observed in the map such as the construction of the E311 road in 2017, Mafraq-Ghweifat highway in 2013 in addition to the increase in the Suburb in Outskirts of Abu Dhabi in 2013 (shown in purple box).

The NTLs have also been analyzed in terms of the seasonal variability and the urbanization structure. Thus, the NTLs images have been classified into winter and summer and the average NTLs was calculated accordingly (Figure 5a-b). It is interesting to see that the total NTLs is lower in the summer season compared to the winter season. This could be attributed to the longer day time during summer compared to the winter. In addition, the movement of population and the less consumption of light during the summer holidays could contribute to this decrease. The seasonal change is highest in the cities of Al-Ain and Al-Dhafra compared to the other cities like Abu Dhabi and Shariah which have shown less change. In regard to the change in the NTLs and the land covers (Figure 6a-b), it can be observed that the industrial regions have developed more NTLs coverage especially over the years between 2016 and 2021 as there is an average increase of 25% of surface NTLs in an annual basis compared to 2012.

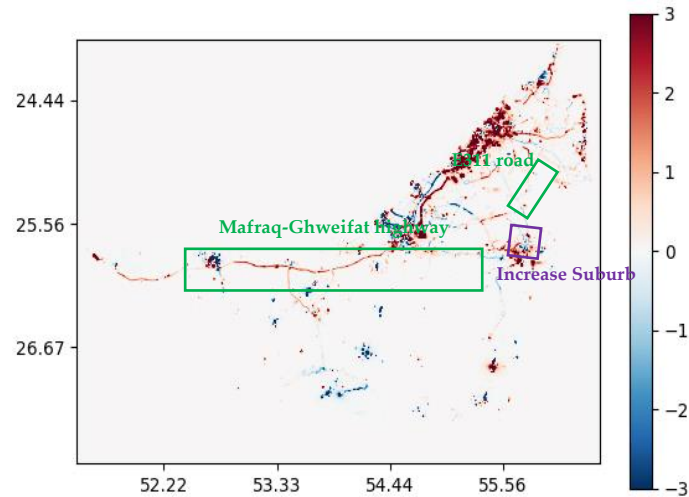

**Figure 4.** Change in spatial trend of NPP-NTLs from 2012 to 2021. Red color indicates positive change, whereas blue color indicates negative change.

In contrast, the NTLs around the farms and agriculture areas are found to be less than that in 2012 in most of the years except for 2019, 2015 and 2021. As for the urban areas, there is a gradual increase in the NTLs as well but to lesser extent compared to the industrial regions. This means that the industrial development in the UAE required high usage of light in the night, compared to the urban regions. This is also seen by the development of the roads over the years.

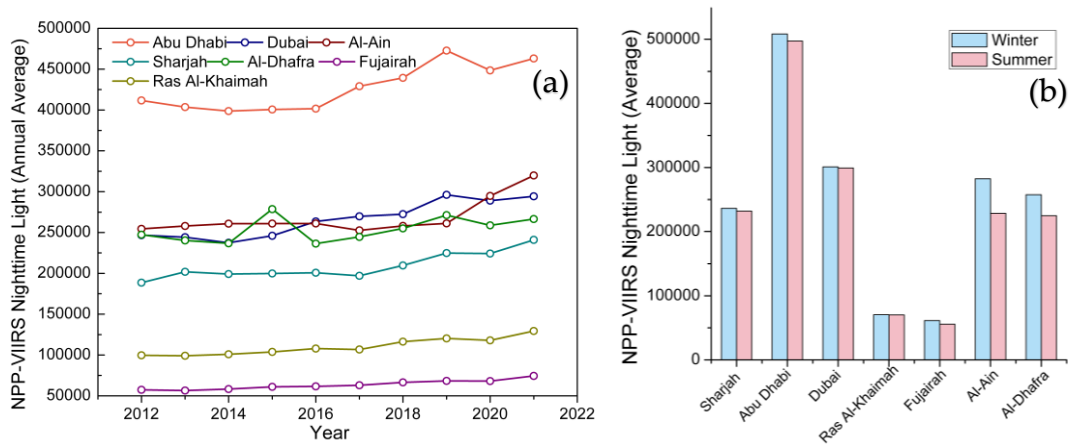

**Figure 5.** (a) Annual changes of NPP-VIIRS Nighttime Light between 2012 and 2021 over seven different cities; b) the average seasonal NPP-VIIRS Nighttime Light over the seven cities in winter and summer.

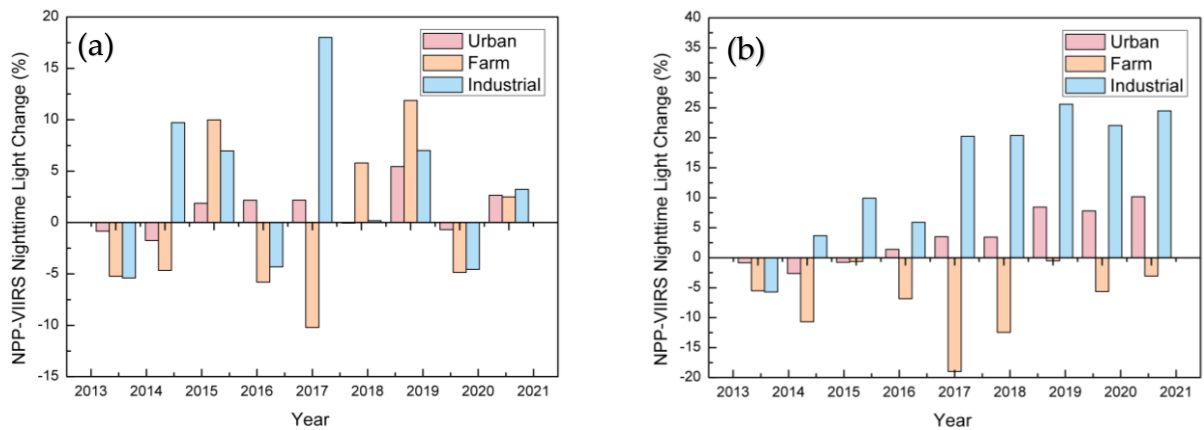

**Figure 6.** NPP-VIIRS Nighttime Light: a) annual changes between 2012 and 2021 in three different land covers of Urban, Farm and industrial areas; b) changes compared to 2012 as a reference year.

*3.3. Correlations between NTLs and $CO_2$, electricity and population*

The daily NTLs values have been averaged into annual average values for the years between 2011 to 2021. The NTLs then were used to find the correlation between them and the socioeconomic variables including the population, $CO_2$ emissions and electrical consumption in the UAE based on the OLS approach. The NTLs are found to be highly correlated with the population and a significant linear model has been developed with an $R^2$ of 0.94 as shown in Figure 7a. The data points are precise and are uniformly distributed along the regression line. Good linear fits were also obtained for the individual cities including the best fit obtained using the data of Dubai city with $R^2$ of 0.855 (Figure 7b). In addition, acceptable linear fits have been obtained for the city of Abu Dhabi (Figure 7c) and Sharjah (Figure 7d) with $R^2$ around 0.7. However, the city of Al-Ain has shown lower correlation and thus weaker linear model is derived with $R^2$ of 0.4 (Figure 7e). This could be explained by the rural nature of this city as a low-electrified region with EC less than $1.1 \times 10^4$ GWh. These results confirm the capability of using the NTLs as indication for the urban related parameters of population and urbanization in urban and rural regions.

As for the correlation between the NTLs and carbon footprint, the total annual NTLs has been used to find the best fit model that can estimate the $CO_2$ emissions. Thus, the results have shown that the NTLs could be highly linearly correlated with the total $CO_2$ emissions in the UAE with $R^2$ around 0.8. This is considered as a good linear fit given the low number of the data points. To further investigate the relation between NTLs and $CO_2$, the atmospheric model data of mol $CO_2$/mol dry air has been extracted over the UAE over the same time period as shown in Figure 8. As shown in this figure, there is a positive correlation between NTLs and $CO_2$ concentration with $R^2$ of 0.5. Similarly, NTLs are found to be correlated with $CO_2$/capita with $R^2$ of 0.6 based on low number of the data points that are distributed along the linear regression line. The linear regression coefficients are shown in Figures 7f-g.

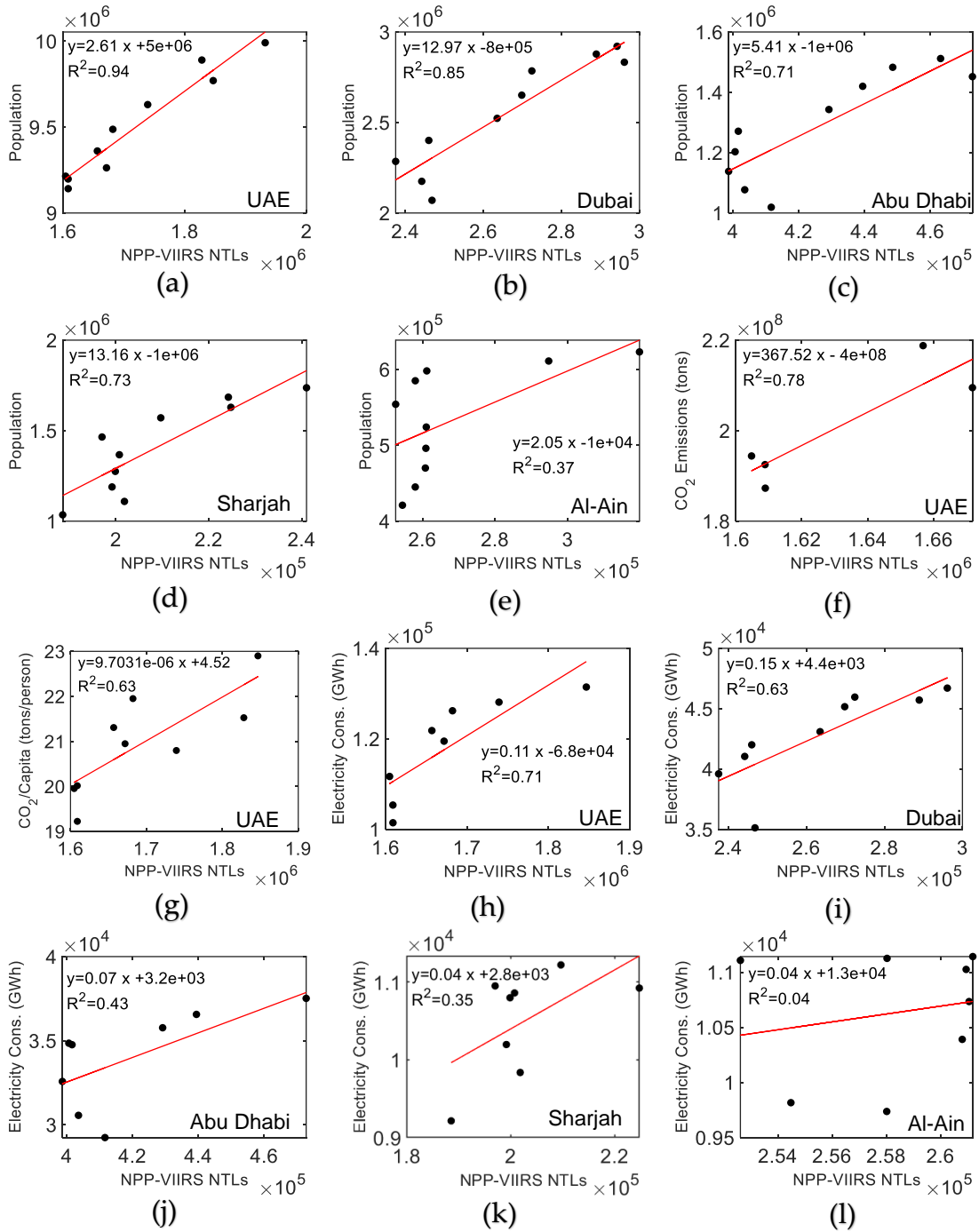

**Figure 7.** Correlation between the NPP-VIIRS Nighttime Light and (a-e) population, (f) CO2 emissions in the UAE, (g) $CO_2$/capita and (h-l) Electricity in the UAE and different cities between 2012 and 2021.

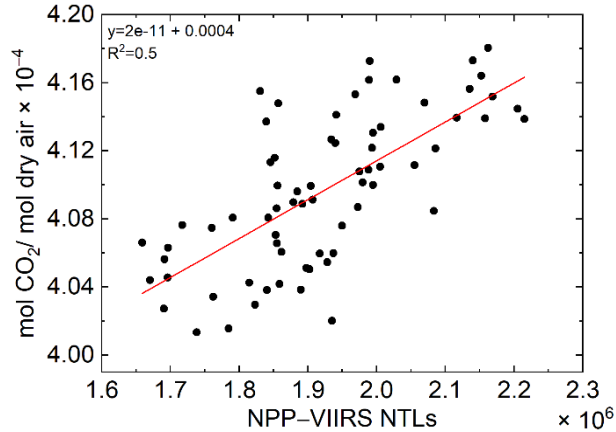

**Figure 8.** Correlation between the NPP-VIIRS Nighttime Light and $CO_2$ in mol $CO_2$/mol dry air extracted from the for the UAE.

On the other hand, NTLs could produce good linear correlations with the EC values but in the UAE and Dubai with $R^2$ more than 0.6. In the cities of Abu Dhabi and Sharjah the best fits are with $R^2$ ranging from 0.3 to 0.45. Whereas, in the Al-Ain and Al-Dhafra, the correlations are weak with $R^2$ less than 0.2. Consequently, this prove that the EC could not be predicted directly by the NTL. One reason for this inconsistency is the limited number of data, and therefore more data could be used to investigate the relation between the NTL and EC (Figures 7h-l).

4. Conclusions

The correlation between the NTL and socioeconomic variables has demonstrated that remote sensing technology can be used to predict population density and carbon footprint. The results show that in the urban regions, there are high correlations between NTLs and the urbanization parameters including population, $CO_2$ emissions and electricity consumption with $R^2$ exceeding 0.6. However, in the rural regions where low-electrified regions are, the correlation between NTLs and EC is low (<0.4). In these rural regions, the EC doesn't exceed $1.1 \times 10^4$ GWh and the annual changes are low of 0.2 GWh as seen in the regions of Al-Ain and Sharjah.

Utilizing the NTLs data since 2012 can be used to assist decision makers in the areas of urban density and environmental quality. Increasing urban residential density will have a significant impact on local emissions and will be most effective at low densities. For example, in Abu Dhabi, residential areas were developed in the last decade, thus the emissions did not increase significantly in comparison to Dubai. In response to the increase in residential area, there has been a decrease in the utilization of public and private transportation. Thus, the decision makers may use the NTLs for studying the development of population density and the possibility of distributing the residents in close quarters to reduce the use of automobiles. In fact, the COVID-19 pandemic has approved that the decrease in transportation and working remotely could decrease the $CO_2$ emissions in several countries in the world [19].

**Competing interests**

The authors have no competing interests to declare that are relevant to the content of this article.

**References**

1. Kumar, P.; Morawska, L.; Martani, C.; Biskos, G.; Neophytou, M.; Di Sabatino, S.; Bell, M.; Norford, L.; Britter, R. The Rise of Low-Cost Sensing for Managing Air Pollution in Cities. *Environ. Int.* **2015**, *75*, 199–205.
2. Ahmad, M.W.; Mourshed, M.; Mundow, D.; Sisinni, M.; Rezgui, Y. Building Energy Metering and Environmental Monitoring–A State-of-the-Art Review and Directions for Future Research. *Energy Build.* **2016**, *120*, 85–102.
3. Ottelin, J.; Ala-Mantila, S.; Heinonen, J.; Wiedmann, T.; Clarke, J.; Junnila, S. What Can We Learn from Consumption-Based Carbon Footprints at Different Spatial Scales? Review of Policy Implications. *Environ. Res. Lett.* **2019**, *14*, 93001.
4. Zhang, Z.; Chen, Y.H.; Wang, C.M. Can $CO_2$ Emission Reduction and Economic Growth Be Compatible? Evidence from China. Front. *Energy Res* **2021**, *9*, 315.
5. Moomaw, W.R.; Law, B.E.; Goetz, S.J. Focus on the Role of Forests and Soils in Meeting Climate Change Mitigation Goals: Summary. *Environ. Res. Lett.* **2020**, *15*, 45009.
6. Scipioni, A.; Manzardo, A.; Mazzi, A.; Mastrobuono, M. Monitoring the Carbon Footprint of Products: A Methodological Proposal. *J. Clean. Prod.* **2012**, *36*, 94–101.
7. Wheeler, S.M.; Jones, C.M.; Kammen, D.M. Carbon Footprint Planning: Quantifying Local and State Mitigation Opportunities for 700 California Cities. *Urban Plan.* **2018**, *3*, 35–51.
8. Koide, R.; Kojima, S.; Nansai, K.; Lettenmeier, M.; Asakawa, K.; Liu, C.; Murakami, S. Exploring Carbon Footprint Reduction Pathways through Urban Lifestyle Changes: A Practical Approach Applied to Japanese Cities. *Environ. Res. Lett.* **2021**, *16*, 84001.
9. Lin, J.; Shi, W. Statistical Correlation between Monthly Electric Power Consumption and VIIRS Nighttime Light. *ISPRS Int. J. Geo-Information* **2020**, *9*, 32.
10. Shi, K.; Yu, B.; Huang, Y.; Hu, Y.; Yin, B.; Chen, Z.; Chen, L.; Wu, J. Evaluating the Ability of NPP-VIIRS Nighttime Light Data to Estimate the Gross Domestic Product and the Electric Power Consumption of China at Multiple Scales: A Comparison with DMSP-OLS Data. *Remote Sens.* **2014**, *6*, 1705–1724.
11. Chen, H.; Zhang, X.; Wu, R.; Cai, T. Revisiting the Environmental Kuznets Curve for City-Level $CO_2$ Emissions: Based on Corrected NPP-VIIRS Nighttime Light Data in China. *J. Clean. Prod.* **2020**, *268*, 121575.
12. Reiche, D. Renewable Energy Policies in the Gulf Countries: A Case Study of the Carbon-Neutral "Masdar City" in Abu Dhabi. *Energy Policy* **2010**, *38*, 378–382.
13. Sbia, R.; Shahbaz, M.; Ozturk, I. Economic Growth, Financial Development, Urbanisation and Electricity Consumption Nexus in UAE. *Econ. Res. istraživanja* **2017**, *30*, 527–549.
14. Worldometer United Arab Emirates CO2 Emissions - Worldometer Available online: https://www.worldometers.info/co2-emissions/united-arab-emirates-co2-emissions/ (accessed on 20 August 2022).
15. De Jong, M.; Hoppe, T.; Noori, N. City Branding, Sustainable Urban Development and the Rentier State. How Do Qatar, Abu Dhabi and Dubai Present Themselves in the Age of Post Oil and Global Warming? *Energies* **2019**, *12*, 1657.

16. Elvidge, C.D.; Zhizhin, M.; Baugh, K.; Hsu, F.-C.; Ghosh, T. Methods for Global Survey of Natural Gas Flaring from Visible Infrared Imaging Radiometer Suite Data. *Energies* **2015**, *9*, 14.
17. Kendall, M.G. A New Measure of Rank Correlation. *Biometrika* **1938**, doi:10.2307/2332226.
18. Yue, S.; Wang, C.Y. Applicability of Prewhitening to Eliminate the Influence of Serial Correlation on the Mann-Kendall Test. *Water Resour. Res.* **2002**, *38*, 1–4.
19. Gocic, M.; Trajkovic, S. Analysis of Changes in Meteorological Variables Using Mann-Kendall and Sen's Slope Estimator Statistical Tests in Serbia. *Glob. Planet. Change* **2013**, *100*, 172–182.
20. Al Shehhi, M.R.; Abdul Samad, Y. Effects Of The Covid-19 Pandemic On The Oceans. *Remote Sens. Lett.* **2021**, *12*, 325–334, doi:10.1080/2150704X.2021.1880658.